\documentclass[aps,pra,reprint,amsmath,superscriptaddress,amssymb,showpacs,nofootinbib,onecolumn]{revtex4-2}
\usepackage{amsmath,braket,amssymb}
\usepackage{multirow}
\usepackage[final]{graphicx}
\usepackage{subfigure}
\usepackage[english]{babel}
\usepackage[utf8]{inputenc}
\usepackage{mathtools}
\usepackage{empheq}
\usepackage{tensor}
\usepackage{cancel}
\usepackage{enumitem}
\usepackage{simplewick}

\usepackage{enumitem}
\usepackage{slashed}
\usepackage{graphicx}
\usepackage{xcolor}

\colorlet{Green}{black!30!green}

\definecolor{THc}{rgb}{0.9,0.3,0.2}

\usepackage{tikz}
\usetikzlibrary{calc,fadings,decorations.pathreplacing,shapes,shapes.multipart,arrows,shapes.misc,intersections,positioning,patterns}
\tikzset{arrow data/.style 2 args={%
		decoration={%
			markings,
			mark=at position #1 with \arrow{#2}},
		postaction=decorate}
}
\usepackage{bm}
\usepackage[colorlinks=true,citecolor=blue,linkcolor=blue,urlcolor=blue]{hyperref}
\usepackage{dsfont}
\usepackage{changepage}
\usepackage{array}
\usepackage{soul}
\usepackage[capitalise]{cleveref}
\crefname{section}{Sec.}{Secs.}
\Crefname{section}{Sec.}{Secs.}
\usepackage{xifthen}
\usepackage{xargs}
\usepackage{hhline}
\usepackage{pifont}

\usepackage{amsthm}
\theoremstyle{definition}

\theoremstyle{plain}

\usepackage{bm}

\graphicspath{{./}{./figures/}}

\newcommand{\bit}{\begin{itemize}}
	\newcommand{\eit}{\end{itemize}}

\renewcommand{\>}{\right\rangle}
\newcommand{\<}{\left\langle}
\newcommand{\ba}{\begin{align}}
	\newcommand{\ea}{\end{align}}
\newcommand{\be}{\begin{equation}}
	\newcommand{\ee}{\end{equation}}
\newcommand{\bi}{\begin{itemize}}
	\newcommand{\ei}{\end{itemize}}

\DeclareMathAlphabet{\mymathbb}{U}{BOONDOX-ds}{m}{n}

\usepackage{physics}

\usepackage{braket}
\usepackage[normalem]{ulem}

\begin{document}
	\date{\today}

	\newcommand{\bbra}[1]{\<\< #1 \right|\right.}
	\newcommand{\kket}[1]{\left.\left| #1 \>\>}
	\newcommand{\bbrakket}[1]{\< \Braket{#1} \>}
	\newcommand{\pll}{\parallel}
	\newcommand{\nn}{\nonumber}
	\newcommand{\transp}{\text{transp.}}
	\newcommand{\nor}{z_{J,H}}
	
	\newcommand{\hL}{\hat{L}}
	\newcommand{\hR}{\hat{R}}
	\newcommand{\hQ}{\hat{Q}}

    	\newcommand{\hh}{\widetilde{h}}

	\title{The Hilbert-space structure of free fermions in disguise}

\begin{abstract}
Free fermions in disguise (FFD) Hamiltonians describe spin chains which can be mapped to free fermions, but not via a Jordan-Wigner transformation. Although the mapping gives access to the full Hamiltonian spectrum, the computation of spin correlation functions is generally hard. Indeed, the dictionary between states in the spin and free-fermion Hilbert spaces is highly non-trivial, due to the non-linear and non-local nature of the mapping, as well as the exponential degeneracy of the Hamiltonian eigenspaces. In this work, we  provide a series of results characterizing the Hilbert space associated to FFD Hamiltonians. We focus on the original model introduced by Paul Fendley and show that the corresponding Hilbert space admits the exact factorization $\mathcal{H}=\mathcal{H}_F\otimes \mathcal{H}_D$, where $\mathcal{H}_F$ hosts the fermionic operators, while $\mathcal{H}_D$ accounts for the exponential degeneracy of the energy eigenspaces. By constructing a family of spin operators generating the operator algebra supported on $\mathcal{H}_D$, we further show that $\mathcal{H}_D=\mathcal{H}_{F'}\otimes \mathcal{H}_{\widetilde{D}}$, where $\mathcal{H}_{F'}$ hosts ancillary free fermions in disguise, while $\mathcal{H}_{\widetilde{D}}$ is generated by the common eigenstates of an extensive set of commuting Pauli strings. Our construction allows us to fully resolve the exponential degeneracy of all Hamiltonian eigenspaces and is expected to have implications for the computation of spin correlation functions, both in and out of equilibrium.
\end{abstract}

\author{Eric Vernier}
\affiliation{Laboratoire de Probabilités, Statistique et Modélisation CNRS, Université Paris Cité, Sorbonne Université Paris, France}

\author{Lorenzo Piroli}
\affiliation{Dipartimento di Fisica e Astronomia, Universit\`a di Bologna and INFN, Sezione di Bologna, via Irnerio 46, I-40126 Bologna, Italy}

\maketitle

	\section{Introduction}
 \label{sec:intro}
It has been almost a hundred years since the Jordan-Wigner transformation (JW) was introduced in theoretical physics~\cite{jordan1993}. Today, this transformation remains an invaluable tool, allowing us to map one-dimensional spin systems into fermionic ones, with several generalizations being proposed over the years~\cite{fradkin1989jordan, wigner1991, huerta1993bose, batista2001generalized, verstraete2005mapping, kitaev2006anyons, nussinov2012arbitrary, chen2018exact, backens2019jordan, tantivasadakarn2020jordan, chapman2020characterization, minami2016solvable, minami2017infinite, yanagihara2020exact, ogura2020geometric, wang2025particle}. Importantly, the JW transformation and its generalizations make it easy to identify and construct exactly-solvable spin chains, which are the ones that are mapped to non-interacting fermionic Hamiltonians~\cite{franchini2017introduction}.

Given the long history of the JW transformation, it is perhaps surprising that only recently new types of spin chains that can be mapped to free fermions were discovered. In 2019, Paul Fendley constructed a model which is quartic in the JW fermions, and still quadratic in terms of suitably defined fermionic creation and annihilation operators~\cite{fendley2019free} (see Refs.~\cite{fendley2007cooper,de2016integrable,feher2019curious} for earlier work in this direction). Remarkably, it was proven that the spin chain found by Fendley cannot be diagonalized by any generalized JW transformation~\cite{elman2021free}, making the model fundamentally different from the previously known ones. This model and later generalizations~\cite{alcaraz2020free,alcaraz2020integrable,fendley2024free,fukai2025quantum,fukai2025free,chapman2023unified} are now identified with the name of free fermions in disguise (FFD).

The mapping to free fermions in FFD models gives access to the full Hamiltonian spectrum. Still, the computation of spin correlation functions is generally hard, because the dictionary between states in the spin and free-fermion Hilbert spaces is highly non-trivial. Indeed, on the one hand the FFD mapping is non-linear and non-local; on the other hand, each Hamiltonian eigenspace is known to be exponentially degenerate~\cite{fendley2019free}, so that fermionic Fock states cannot be uniquely associated with spin states. As a consequence, the problem of computing correlation functions in FFD models remains essentially open, {despite some recent progress in addressing the ``inverse problem" of expressing local spin operators in terms of fermionic ones~\cite{vona2014exact}. In some respect, the current situation is similar to that of interacting integrable models~\cite{korepin1997quantum,essler2005one}: while the Bethe ansatz gives direct access to the Hamiltonian spectrum, computing correlation functions in arbitrary eigenstates is hard, and analytic results exist only in a few special cases~\cite{kitanine1999form,kitanine2000correlation,boos2007factorization,caux2007one,trippe2010short,pozsgay2011local, negro2013one, mestyan2014short,piroli2015exact, piroli2016multiparticle, pozsgay2017excited, bastianello2018exact,bastianello2018sinh}.

The picture is very different for JW free fermions, where a number of simplifications occur. First, the JW transformation maps local bilinear fermionic operators into local spin operators. Second, it is often easy to characterize the family of spin states that map to the fermionic Gaussian states\footnote{We recall that Gaussian states are defined as those satisfying the Wick theorem~\cite{bravyi2004lagrangian,surace2022fermionic}.}. These facts have consequences for the complexity of the JW solvable models. For instance, the dynamics of spin correlation functions can be computed either analytically or numerically efficiently in many cases of interest~\cite{terhal2002classical,vanDenNest2011simulating,brod2016efficient}, and especially in the case of quantum quenches~\cite{calabrese2006time,fagotti2008evolution,calabrese2011quantum,essler2016quench}.  In addition, the eigenstate wave functions are simple compared to generic Hamiltonians. In particular, all eigenstates can be prepared efficiently by a quantum computer, as they can be obtained as the output of a quantum circuit whose depth scales only linearly in the system size~\cite{verstraete2009quantum,ferris2014fourier,kivlichan2018quantum,jiang2018quantum,hackl2018circuit}. It is natural to ask whether the same conclusions can be drawn for FFD models. Can the system dynamics be simulated on a classical computer? Can all eigenstates be prepared efficiently on a quantum computer?~\cite{lutz2025adiabatic} These questions are non-trivial, due to the complicated structure of the dictionary between spin and fermion degrees of freedom. 

In this work, we take a step towards answering these questions, by characterizing the Hilbert space associated to FFD Hamiltonians. For concreteness, we will focus on the original model introduced by Paul Fendley~\cite{fendley2019free}, although similar constructions hold for more general FFD Hamiltonians. We show that the corresponding Hilbert space admits the exact factorization $\mathcal{H}=\mathcal{H}_F\otimes \mathcal{H}_D$, where $\mathcal{H}_F$ hosts the fermionic operators, while $\mathcal{H}_D$ accounts for the exponential degeneracy of the energy eigenspaces. By constructing a family of spin operators generating the operator algebra supported on $\mathcal{H}_D$, we further show that $\mathcal{H}_D=\mathcal{H}_{F'}\otimes \mathcal{H}_{\widetilde{D}}$, where $\mathcal{H}_{F'}$ hosts ancillary free fermions in disguise, while $\mathcal{H}_{\widetilde{D}}$ is generated by the common eigenstates of an extensive set of commuting Pauli strings. Our construction allows us to fully resolve the exponential degeneracy of all Hamiltonian eigenspaces and is expected to have implications for the computation of spin correlation functions, both in and out of equilibrium.

This work is organized as follows. We begin in Sec.~\ref{sec:the_model} by recalling the model introduced by Paul Fendley (Sec.~\ref{sec:hamiltonian_def}) and its exact solution in terms of free fermions (Sec.~\ref{eq:exact_solution}). Our results are contained in Secs.~\ref{sec:HS_structure}, \ref{sec:tilde_h}, and~\ref{sec:ancillary_fermions}, and summarized in Sec~\ref{sec:result_overview}. In Sec.~\ref{sec:HS_structure} we show that the Hilbert space factorizes as $\mathcal{H}=\mathcal{H}_F\otimes \mathcal{H}_D$, where $\mathcal{H}_F$ supports the fermionic operators. The structure of $\mathcal{H}_D$ is further resolved in the later sections, where we show that $\mathcal{H}_D=\mathcal{H}_{F'}\otimes \mathcal{H}_{\widetilde{D}}$ and characterize the two factors individually. In particular, in Sec.~\ref{sec:tilde_h} we show that $\mathcal{H}_{\widetilde{D}}$ is a ``representation-dependent'' subspace, generated by the common eigenstates of an extensive set of commuting Pauli strings, while in Sec.~\ref{sec:ancillary_fermions} we show that the space of operators supported on $\mathcal{H}_{F'}$ is generated by an operator algebra of ancillary free fermion in disguise. Finally, Sec.~\ref{sec:outlook} contains our conclusions and an outlook for future work.

\section{The Model and its Hilbert-space structure}
\label{sec:the_model}
In this section, we review the FFD construction and introduce the model studied in this work.

\subsection{The Hamiltonian}
\label{sec:hamiltonian_def}
We first introduce the FFD Hamiltonian presented in Ref.~\cite{fendley2019free}. The construction starts with an abstract operator algebra defined in terms of a set of generators $\{h_m\}_{m=1}^{M}$. They satisfy the commutation relations
\begin{align}
(h_m)^2 &= 1 \,, \nonumber\\
\{h_m,h_{m+1}\} &= \{h_m,h_{m+2}\}=0 \,,\nonumber\\
[h_m,h_{l}]&=0\,, \qquad |m-l|>2\nonumber\,.
\end{align}
Different representations of this algebra give rise to different models. In this work, we focus on the original representation considered in Ref.~\cite{fendley2019free}, which is associated with a one-dimensional chain of $M$ spins. The Hilbert space of the model is $\mathcal{H}= (\mathbb{C}^2)^{\otimes M}$, while
\be
h_m = Z_{m-2} Z_{m-1} X_m  \,,\qquad m=1\ldots M \,.
\label{representation}
\ee 
Here, $X_j$ and $Z_j$ with $j\in\{1, \ldots , M\}$ are the Pauli matrices acting on spin $j$, while we use the convention $Z_{-2}=Z_{-1}=1$. As mentioned, although we focus on the representation in Eq.~\eqref{representation}, our constructions are expected to hold for more general representations of the generators $h_m$ (more specifically our splitting of the Hilbert space will make clear which part of the construction depends on the representation, and which part does not).

With these notations, the FFD Hamiltonian is defined by
\be
\label{eq:FFD_Hamiltonian}
H = \sum_{m=1}^{M} b_m h_m \,,
\ee 
where $b_m\in \mathbb{R}$ are arbitrary real numbers parametrizing the model (note that the Hamiltonian~\eqref{eq:FFD_Hamiltonian} corresponds to open boundary conditions). In the next subsection, we review the exact solution of the Hamiltonian~\eqref{eq:FFD_Hamiltonian}. We will focus on the aspects that are directly relevant to our work, and refer to Ref.~\cite{fendley2019free} for more detail.

\subsection{Exact solution}
\label{eq:exact_solution}

The diagonalization of the Hamiltonian~\eqref{eq:FFD_Hamiltonian} is achieved by constructing suitable fermionic creation and annihilation operators. 
A key object is the transfer matrix $T_M(u)$, an operator acting on $\mathcal{H}$ that depends on a complex parameter $u$, called the spectral parameter. The transfer matrix can be constructed as the generating function of the conserved quantities of the model \cite{fendley2019free}, in a fashion which is reminiscent of Bethe-ansatz solvable models~\cite{korepin1997quantum}. 


As an important property, transfer matrices at different spectral parameters commute, $[T_M(u),T_M(u^\prime)]=0$, and the Hamiltonian can obtained as the logarithmic derivative 
\begin{equation}
\label{eq:trace_formula}
H=-\frac{d}{du}\ln T_M(u)\big|_{u=0}\,.
\end{equation}
The transfer matrix can be written entirely in terms of the generators $h_m$ and the real numbers $\{b_m\}$. Exploiting the algebra of the former, one can write explicitly the operator $T_M^{-1}(u)$ involved in the logarithmic derivative. In particular, one shows that $T_M(u) T_M(-u)$ is proportional to the identity, namely 
\begin{equation}
\label{eq:inversion_formula}
T_M(u) T_M(-u)
=
P_M\left(u^2\right)\,,
\end{equation}
so that $T_M^{-1}(u)\propto T_M(-u)$. Here $P_M(u^2)$ is a polynomial of degree $S=[(M+2)/3]$ in $u^2$  ($[\cdot]$ denotes the integer part), whose zeroes $\{u_k^2\}_{k=1,\ldots S}$ uniquely determine the energy spectrum of $H$: denoting $\epsilon_k=1/u_k$, there are $2^S$ distinct energy eigenvalues of the form 
\be 
E = \pm \epsilon_1 \pm \epsilon_2 \pm \ldots \pm \epsilon_S \,,
\ee 
where all signs can be chosen independently. Each eigenvalue corresponds to a degenerate subspace of dimension $2^{M-S}$.

The transfer matrix also allows us to express the associated raising and lowering operators. The construction requires introducing an extra operator $\chi$ that anti-commutes with $h_M$, but which commutes with the rest of the generators. Consistent with the representation~\eqref{representation}, we choose 
\begin{equation}
\label{eq:extra_operator}
\chi=Z_M\,.
\end{equation}
As shown in \cite{fendley2019free}, the creation/annihilation operators associated with the energy $\epsilon_k$ then take the form
\be 
\Psi_k \propto T_M(\epsilon_k) ~\chi ~ T_M(-\epsilon_k)\,.
\label{eq:psikdef}
\ee 
It can be shown that $\Psi_k$ are indeed fermionic operators, satisfying
\begin{equation}
\{\Psi_k,\Psi^\dagger_q\}=\delta_{q,k}\,,
\end{equation}
and $\Psi_k^2=0$. 

Finally, the FFD Hamiltonian can be written in diagonal form in terms of the fermionic operators~\eqref{eq:psikdef} as
\begin{equation}
    H=\sum_{k=1}^S \epsilon_k\left[\Psi_k, \Psi_{-k}\right]\,,
\end{equation}
where we used the notation $\Psi_{-k}=\Psi^\dagger_{k}$ for $k\geq 1$.

\subsection{The Hilbert-space structure}
\label{sec:HS_structure}

In this section, we use the fermionic operators \eqref{eq:psikdef} to show that the FFD Hilbert space can be decomposed as $\mathcal{H}=\mathcal{H}_F\otimes \mathcal{H}_D$. 
Here, $\mathcal{H}_F$ is a space of dimension $2^S$ that the fermionic operators are supported on, while $\mathcal{H}_D$ accounts for the exponentially large degeneracies observed in the spectrum of $H$. This simple observation is our first result and the basis for the subsequent analysis presented in the rest of this work, cf.~Sec.~\ref{sec:result_overview}

We begin by defining the subspace which is annihilated by all annihilation operators $\Psi_k$,
\begin{equation}
    \mathcal{K}=\{\ket{v}\in \mathcal{H}\ :\  \Psi_k\ket{v}=0\ ~ \forall k=1\,,\ldots S\}\,.
\end{equation}
Clearly, $\mathcal{K}$ coincides with the degenerate space associated with the ground-state energy, and has dimension ${\rm dim}(\mathcal{K})=2^{M-S}$, cf Sec.~\ref{eq:exact_solution}. We denote by $\{\ket{w_j}\}_{j=1}^{2^{M-S}}$ an orthogonal basis of $\mathcal{K}$. Next, we define
\begin{equation}
\label{eq:basis_states}
    \ket{f_{j,\alpha}}=(\Psi^\dagger_{1})^{\alpha_1} (\Psi^\dagger_{2})^{\alpha_2}\cdots (\Psi^\dagger_{S})^{\alpha_S}\ket{w_j}\,,
\end{equation}
where $\alpha_j=0,1$, and $(\alpha_1\alpha_2,\ldots \alpha_S)$ is the binary representation of the integer $\alpha\in [0,\ldots 2^S-1]$. Because of the properties of the fermionic operators, it is immediate to see that $\ket{f_{j,\alpha}}$ are mutually orthogonal, and thus form a basis for the Hilbert space $\mathcal{H}$.

The basis~\eqref{eq:basis_states} allows us to show the previously announced Hilbert-space factorization $\mathcal{H}=\mathcal{H}_F\otimes \mathcal{H}_D$. Formally, we do this by introducing the map
\begin{align}
    \mathcal{\varphi}:\quad \mathcal{H}\quad &\to \mathcal{H}_F\otimes \mathcal{H}_D\nonumber\\
        \ket{f_{j,\alpha}}&\mapsto \ket{\alpha}\otimes \ket{j}\,.
\end{align}
Here, $\mathcal{H}_F$ is a Hilbert space generated by the fermionic Fock states
\begin{equation}\label{eq:fermionic_fock_states}
 \ket{\alpha}=(c^\dagger_{1})^{\alpha_1} (c^\dagger_{2})^{\alpha_2}\cdots (c^\dagger_{S})^{\alpha_S}|\Omega\rangle\,,
\end{equation}
where $\{c_j^\dagger\}$ are canonical fermionic creation operators, $\ket{\Omega}$ is the corresponding vacuum state, while $\mathcal{H}_D$ is generated by the basis vectors $\{\ket{j}\}_{j=1}^{2^{M-S}}$. Note that the order of the operators $\Psi_j$ and $c_j$ in the definitions~\eqref{eq:basis_states} and~\eqref{eq:fermionic_fock_states} is fixed. It is then easy to show that $\varphi$ is an isomorphism preserving the Hilbert-Schmidt scalar product. 

Next, we claim that the isomorphism $\varphi$ maps the operators $\Psi_k$ into operators supported on $\mathcal{H}_F$, namely $\varphi(\Psi_k)=c_k\otimes \openone $. Indeed, by definition of $\varphi$,
\begin{align}
\bra{\alpha}\bra{j} \varphi(\Psi_k) \ket{\beta}\ket{k}&=\langle f_{j,\alpha}|\Psi_k|f_{k,\beta}\rangle\nonumber\\
&=\delta_{j,k}\langle \alpha|c_k|\beta\rangle\,,
\end{align}
so that all fermionic operators are supported on $\mathcal{H}_F$. Finally, since the Hamiltonian is written explicitly in terms of the fermionic operators, it follows that (with a slight abuse of notation) $H=H\otimes \openone$. Therefore, $\mathcal{H}_D$ accounts for the exponential degeneracy of each Hamiltonian eigenstate.

\section{Overview of our results}
\label{sec:result_overview}
We have shown that the Hilbert space admits the factorization $\mathcal{H}=\mathcal{H}_F\otimes\mathcal{H}_{D}$. In the rest of this work, we will further refine the structure of $\mathcal{H}_D$. Before proceeding, it is useful to provide an overview of our main results, which we report in this section. 

In essence, we will show that $\mathcal{H}_D$ can be further factorized as $\mathcal{H}_D=\mathcal{H}_{\widetilde{D}} \otimes \mathcal{H}_{F'}$, we will exhibit an explicit orthonormal basis for $\mathcal{H}_{\widetilde{D}}$ and $\mathcal{H}_{F'}$, and characterize the operator algebras supported on $\mathcal{H}_{\widetilde{D}}$ and $\mathcal{H}_{F'}$, respectively. In more detail, our results can be summarized as follows: 
\begin{itemize} 
\item We will show that $\mathcal{H}_{\widetilde{D}}$ supports the action of all operators that commute with all the generators $h_m$, as well as with the boundary mode $\chi$ (and therefore, with all fermionic operators $\{\Psi_k\}, \{\Psi^\dagger_k\}$). We will denote the algebra of such operators by $\mathcal{A}_{\widetilde{D}}$. As it will be clear from our analysis, $\mathcal{A}_{\widetilde{D}}$ strongly depends on the representation chosen for $h_m$, but not on the values of the couplings $\{b_m\}$ in the Hamiltonian. We will characterize $\mathcal{A}_{\widetilde{D}}$ by exhibiting $2N_P+N_C$ independent generators, consisting of:
\begin{itemize}
    \item $N_P$ pairs of operators $\{(\tilde{X}_m,\tilde{Y}_m)\}_{m=1}^{N_P}$, such that  
\begin{subequations}    
\begin{align}
[\tilde{X}_n,\tilde{X}_{m}]&=[\tilde{Y}_n,\tilde{Y}_{m}]=0 \,, \\
[\tilde{X}_n,\tilde{Y}_{m}] &= 2\delta_{n,m} \,.
\end{align}
\end{subequations}  
    \item $N_C$ central elements commuting with all other elements in the algebra $\mathcal{A}_{\widetilde{D}}$. 
\end{itemize}
The dimension of the auxiliary space $\mathcal{H}_{\widetilde{D}}$ is found to be
\begin{eqnarray}
    {\rm dim}(\mathcal{H}_{\widetilde{D}})=2^{N_P+N_C}\,.
\end{eqnarray}

\item We will show that ${\mathcal{H}}_{F'}$ supports the action of all operators written in terms of the generators $h_m$ that commute with the fermion creation/annihilation operators. We will denote the algebra of such operators by $\mathcal{A}_{F^\prime}$ and we will refer to it as the \emph{ancillary} fermionic algebra (to be distinguished by the fermionic algebra $\mathcal{A}_F$ supported on $\mathcal{H}_F)$. Importantly, the definition of $\mathcal{A}_{F^\prime}$ is independent of the specific representation chosen, because it can be stated entirely in terms of the generators $h_m$ and their algebraic relations. However, we will show that $\mathcal{A}_{F^\prime}$ does depend on the Hamiltonian couplings $\{b_m\}$. Going further, we will show in Sec. \ref{sec:ancillary_fermions}, that $\mathcal{A}_{F^\prime}$ is a fermionic algebra, generated by a new family of ancillary fermion operators
\begin{eqnarray}
    \{\Psi_k^\prime\}_{k=1}^{S'}\,,
\end{eqnarray}
which commute with the original ones
\begin{equation}
[\Psi_j,\Psi'_{k}]=0\,. 
\end{equation}
\end{itemize}
Accordingly, the dimension of the Hilbert space $\mathcal{H}_{F'}$ is found to be
\begin{eqnarray}
    {\rm dim}(\tilde{\mathcal{H}}_{F^{\prime}})=2^{S^{\prime}}\,.
\end{eqnarray}

The dimensions of the subspaces $\mathcal{H}_F$, $\mathcal{H}_{F^\prime}$, and $\mathcal{H}_{\widetilde{D}}$ depend, respectively, on the integers $S$, $S^\prime$, $N_P$ and $N_C$ introduced above. In Table \ref{table:summary}, we summarize their dependence on the system size $M$. In all cases (except for the case $M=6k+3$, where an additional zero mode has to be included to the construction, cf. the end of Sec. \ref{sec:ancillary_fermions}), we can check that the product of the dimension of the subspaces $\mathcal{H}_F$, $\mathcal{H}_{F^\prime}$, and $\mathcal{H}_{\widetilde{D}}$ is equal to the dimension of the full Hilbert space, ${\rm dim}(\mathcal{H})=2^N$.
\begin{table}
\centering
\begin{tabular}{c | c || c |    c || c |}  
\cline{2-5} &  \multicolumn{1}{ c||}{$\mathcal{H}_F$ }&  \multicolumn{2}{c||}{$\mathcal{H}_{\widetilde{D}}$} &  \multicolumn{1}{c|}{$\mathcal{H}_{F'}$}
\\[1ex]
 \hline 
 \multicolumn{1}{|c||}{$M$} & ~$S=\lfloor \frac{M+2}{3} \rfloor$~  & ~~$N_C$~~ & ~~$N_P$~~   &  $~S'=\lfloor \frac{M+2}{6} \rfloor$~  \\[1ex] \hline\hline 
  \multicolumn{1}{|c||}{$6k$} & $2k$ & $1$ & $3k-1$ & $k$ \\
    \multicolumn{1}{|c||}{$6k+1$}  & $2k+1$ & $0$ & $3k$ & $k$\\
  \multicolumn{1}{|c||}{$6k+2$} & $2k+1$ & $1$ & $3k$ & $k$  \\
  \multicolumn{1}{|c||}{$6k+3$} & $2k+1$ & $0$ & $3k+1$ & $k$  \\
  \multicolumn{1}{|c||}{$6k+4$} & $2k+2$ & $0$ & $3k+1$ & $k+1$  \\
  \multicolumn{1}{|c||}{$6k+5$} & $2k+2$ & $1$ & $3k+1$ & $k+1$  \\
\hline
\end{tabular}
\caption{Counting of dimensions associated with each of the subspaces $\mathcal{H}_F$, $\mathcal{H}_D$, $\mathcal{H}_{F'}$. Notations are explained in the main text.}
\label{table:summary}
\end{table}

It is important to compare our results with some constructions found in Ref.~\cite{fendley2019free}. There, a family of operators $E^{(n)}, O^{(n)}$ is presented, commuting with the Hamiltonian and generating an extended supersymmetry (SUSY) algebra. While this symmetry also accounts for the degeneracies of the spectrum, it has a few drawbacks. First, it is not clear what the underlying algebra is. Second, the generators $E^{(n)}, O^{(n)}$ mix with the fermions: namely, they entangle the spaces $\mathcal{H}_F$ and $\mathcal{H}_D$, which we believe makes it difficult to understand the corresponding algebra~\footnote{In our construction, we will also encounter operators $E^{(n)}, O^{(n)}$. Those are defined much similarly to Fendley's, but with the important difference that their relation with the fermionic operators is clearly understood.}.  

\section{Representation-dependent degeneracies: the subspace $\mathcal{H}_{\widetilde{D}}$}
\label{sec:tilde_h}

As pointed out in Ref.~\cite{fendley2019free}, the representation \eqref{representation} allows for a family of operators $\{\widetilde{h}_m\}$ that commute simultaneously with all the $h_m$, and therefore generate a symmetry of the model which does not depend on the Hamiltonian couplings $\{b_m\}$. More generally, such symmetry extends to any type of quantum dynamics defined in terms of the operators $\{h_m\}$, such as the circuits considered in Refs.~\cite{fukai2025quantum,szasz2025construction}. 
The operators $\widetilde{h}_m$ can be chosen as \cite{fendley2019free}
\be 
\widetilde{h}_m = X_m Z_{m+1} Z_{m+2} ,,\qquad m=1\,,\ldots M  \,,
\ee 
with the convention that $Z_{M+1}=Z_{M+2}=1$. 

While $[h_m,\widetilde{h}_n]=0$ for all $m,n \in \{1,\ldots M\}$, only the first $M-1$ operators $\{\widetilde{h}_m\}_{m=1}^{M-1}$ commute with the boundary mode $\chi$, and hence with the physical fermion creation/annihilation operators. These $M-1$ operators generate the symmetry algebra $\mathcal{A}_{\widetilde{D}}$, which is the subject of this section. 

In order to characterize $\mathcal{A}_{\widetilde{D}}$, we first observe that one can get a set of mutually commuting operators out of $\{\hh_m\}_{m=1}^{M-1}$ by selecting one out of every three of them, say $\{\hh_1, \hh_4, \ldots\}$. This set is not yet maximal, as one can still add every other pair of consecutive operators out of the remaining $\hh_m$, say $\{i \hh_2 \hh_3, i \hh_8\hh_9 , \ldots\}$. In total, this results in a set of $\sim M/3+M/6 = M/2$ mutually commuting operators, whose eigenvalues $\pm 1$ account for a $2^{\sim M/2}$-fold degeneracy of the Hamiltonian spectrum.

In the following, we will make this description precise, separating the discussion depending on the value of $M$ modulo $6$. In each case, we will recast the algebra of the $\{\hh_m\}_{m=1}^{M-1}$ in terms of a maximal abelian family $\{\tilde{X}_n\}$, along with a second abelian family $\{\tilde{Y}_n\}$ such that 
\be 
[\tilde{X}_n,\tilde{X}_{m}]=[\tilde{Y}_n,\tilde{Y}_{m}]=0 \,, \qquad [\tilde{X}_n,\tilde{Y}_{m}] = 2\delta_{n,m} \,,
\label{XYcom}
\ee 
each pair $(\tilde{X}_n,\tilde{Y}_n)$ accounting for a two-fold degeneracy of the Hamiltonian spectrum. 
For some values of $M$ modulo $6$, these two families are completed by either one or two additional central operators, commuting with all $\tilde{X}_m$ and $\tilde{Y}_m$. Of these central operators, some belong exclusively to the algebra of the $\hh_m$, while others can also be formulated as elements of the algebra of the $h_m$ (see paragraphs \ref{sec:6kp4} and \ref{sec:6kp5}): for reasons that will be made clear below, the index $N_C$ in Table \ref{table:summary} counts only the former.

\subsection{$M=6k$} 
Let us first consider $M=6k$ for some integer $k$. In this case, a maximal set of mutually commuting operators can be constructed as 
\be 
\{\tilde{X}_n\} = \{\hh_{3l+1}\}_{l=0,\ldots 2k-2} \cup \{  i \hh_{6l+2}\hh_{6l+3}\}_{l=0,\ldots k-1}  \,,
\label{Xk6k}
\ee 
together with the central operator 
\be 
C = \prod_{l=0}^{k-1}i \hh_{6l+2}\hh_{6l+3}\hh_{6l+4} \,.
\label{central6k}
\ee 
For each $\tilde{X}_n$, we can define an operator $\tilde{Y}_{n}$ such that the $\tilde{Y}_n$ commute with one another and with all the $\tilde{X}_m$, but one, see Eq.~\eqref{XYcom}.
The $\tilde{Y}_n$ are constructed as strings of operators $\{\hh_m\}_{m=1}^{M-1}$. For example, it is easy to check that $\tilde{Y}_1 = \hh_3 \hh_5 \hh_9\hh_{11} \ldots \hh_{M-3}\hh_{M-1}$ commutes with all operators \eqref{Xk6k} but the first, as well as with the central operator \eqref{central6k}. The other $\tilde{Y}_n$ are defined analogously. 

The $3k-1$ operators $\tilde{X}_n$, the $3k-1$ operators $\tilde{Y}_n$ and the central operator $C$ generate the full algebra of the $\{\hh_m\}_{m=1}^{M-1}$. They act on a Hilbert space of dimension $2^{3k}$, a basis of which can be defined by specifying the eigenvalues $\pm 1$ of all $\{\tilde{X}_n\}\cup C$, while the $\tilde{Y}_n$ flip those eigenvalues individually. In summary, the corresponding symmetry space has the structure  
\be 
\mathcal{H}_{\widetilde{D}} = (\mathbb{C}^2)^{\otimes 3k-1} \oplus (\mathbb{C}^2)^{\otimes 3k-1} \,,
\ee 
where the direct sum splits the space according to the eigenvalue of $C$, while each individual $\mathbb{C}^2$ carries a representation of one pair $(\tilde{X}_n,\tilde{Y}_n)$, where the other $\tilde{X}_k$, $\tilde{Y}_k$ for $k\neq n$ act trivially.

\subsection{$M=6k+1$}

When $M\ {\rm mod}\ 6=1$, a maximal abelian family can be obtained as 
\be 
\{\tilde{X}_n\} = \{\hh_{3l+1}\}_{l=0,\ldots 2k-1} \cup \{  i \hh_{6l+2}\hh_{6l+3}\}_{l=0,\ldots k-1}  \,,
\label{Xk6kp1}
\ee 
while there is no central operator. As in the previous case, we can construct a family $\{\tilde{Y}_n\}$ obeying \eqref{XYcom}. These $M-1$ operators generate the full algebra of the $\{\hh_m\}_{m=1\ldots M-1}$, and define a symmetry space of the form 
\be 
\mathcal{H}_{\widetilde{D}} = (\mathbb{C}^2)^{\otimes 3k}  \,.
\ee

\subsection{$M=6k+2 $}
\label{sec:6kp2}

When $M\ {\rm mod}\ 6=2$, a maximal set of mutually commuting operators is given by
\be 
\{\tilde{X}_n\} = \{\hh_{3l+1}\}_{l=0,\ldots 2k-1} \cup \{i   \hh_{6l+2}\hh_{6l+3}\}_{l=0,\ldots k-1}  \,,
\label{Xk6kp2}
\ee 
together with the central operator 
\be 
C = \prod_{l=0}^{2k}\hh_{3l+1}\,.
\label{central6kp2}
\ee 
Once again, we can associate to $\{\tilde{X}_n\}$ a family $\{\tilde{Y}_n\}$, obeying \eqref{XYcom}.

In total, we then find $2\times 3k + 1= M-1$ operators, generating the full algebra of the $\{\hh_m\}_{m=1\ldots M-1}$. As in the previous section, those span a symmetry space with structure
\be 
\mathcal{H}_{\widetilde{D}}= (\mathbb{C}^2)^{\otimes 3k}  \oplus 
(\mathbb{C}^2)^{\otimes 3k} \,.
\ee 

\subsection{$M=6k+3 $}

When $M\ {\rm mod}\ 6=3$, a maximal set of mutually commuting operators is
\be 
\{\tilde{X}_n\} = \{\hh_{3l+1}\}_{l=0,\ldots 2k} \cup \{  i \hh_{6l+5}\hh_{6l+6}\}_{l=0,\ldots k-1}  \,,
\label{Xk6kp3}
\ee 
to which we can as usual associate the operators $\{\tilde{Y}_n\}$ satisfying \eqref{XYcom}. These $2\times(3k+1)=M-1$ generate the full algebra of the $\{\hh_m\}_{m=1}^{M-1}$, and define a symmetry space of the form 
\be 
\mathcal{H}_{\widetilde{D}}= (\mathbb{C}^2)^{\otimes 3k+1}\,.
\ee 

\subsection{$M=6k+4 $}
\label{sec:6kp4}

When $M\ {\rm mod}\ 6=4$, a maximal set of mutually commuting operators can be constructed as 
\be 
\{\tilde{X}_n\} = \{\hh_{3l+1}\}_{l=0,\ldots 2k} \cup \{  i \hh_{6l+5}\hh_{6l+6}\}_{l=0,\ldots k-1}  \,,
\label{Xk6kp4}
\ee 
together with the central operator 
\be 
C = \prod_{l=0}^{k}i \hh_{6l+1}\hh_{6l+2}\hh_{6l+3}
= \prod_{l=0}^{k}i h_{6l+1}h_{6l+2}h_{6l+3}
\,.
\label{central6kp4}
\ee 
Again, we can construct a family $\{\tilde{Y}_n\}$, obeying \eqref{XYcom}. We obtain a total of $2\times (3k+1) + 1= M-1$ operators that generate the full algebra of the $\{\hh_m\}_{m=1\ldots M-1}$.
In contrast to the previously encountered cases where the central operator splits the Hilbert space $\mathcal{H}_D$ according to its $\pm 1$ eigenvalue, in the present case we simply write the corresponding symmetry space as 
\be 
\mathcal{H}_{\widetilde{D}}= (\mathbb{C}^2)^{\otimes (3k+1)}  \,.
\ee 
This is because the central generator \eqref{central6kp4} also admits an expression in terms of the generators $h_m$, and therefore is simultaneously a central element of the ancillary free-fermionic algebra to be discussed in the next section.

\subsection{$M=6k+5 $}
\label{sec:6kp5}

Finally, let us consider $M=6k+5$ for some integer $k$. In this case, a maximal set of mutually commuting operators is
\be 
\{\tilde{X}_n\} = \{\hh_{3l+1}\}_{l=0,\ldots 2k} \cup \{  i \hh_{6l+2}\hh_{6l+3}\}_{l=0,\ldots k-1}  \,,
\label{Xk6kp5}
\ee 
together with the central operators 
\begin{subequations}
\begin{align}
C_1 &=   \prod_{l=0}^{k}i \hh_{6l+1}\hh_{6l+2}\hh_{6l+3}\,, \\ 
C_2 &= \prod_{l=0}^{k}i \hh_{6l+2}\hh_{6l+3}\hh_{6l+4} \,.
\end{align}
\label{central6kp5}
\end{subequations}

\noindent Once again, we can construct a family  $\{\tilde{Y}_n\}$ obeying \eqref{XYcom}. Together with \eqref{Xk6kp5} and \eqref{central6kp5}, we thus have a total of $2\times (3k+1) + 2= M-1$ operators, generating the full algebra of the $\{\hh_m\}_{m=1\ldots M-1}$.

In this case, we write the Hilbert space $\mathcal{H}_{\widetilde{D}}$ as
\be 
\mathcal{H}_{\widetilde{D}}= (\mathbb{C}^2)^{\otimes (3k+1)} \oplus (\mathbb{C}^2)^{\otimes (3k+1)}  \,,
\ee 
where the direct sum splits the space according to the eigenvalue of $C_1$. Analogously to what was observed for the case $M=6k+4$, the reason why $C_2$ does not induce further splitting is that it can also be written in terms of the $\{h_m\}$ operators, $C_2 = \prod_{l=0}^{k}h_{6l+2}h_{6l+3}h_{6l+4}$. Therefore, $C_2$ belongs to the algebra of ancilliary FFD presented in the next section.

\subsection{Factoring out the symmetry: mapping to another FFD chain}

\label{sec:twochain}

Having identified the generators of the algebra $\mathcal{A}_{\widetilde{D}}$, it is possible to reduce the size of the FFD Hilbert space (namely, reduce the number of degeneracies) by selecting the eigenvalues of a set of mutually commuting operators in $\mathcal{A}_{\widetilde{D}}$. 

We illustrate this fact by projecting onto the eigenspaces of {\it some} of the symmetry generators, namely $\hh_{3l+1}=X_{3l+1} Z_{3l+2} Z_{3l+3}$. For concreteness we choose to project onto the subspace where they all have eigenvalue $+1$ (a similar discussion holds for any choice of signs). Each block of three spins on sites $3l+1,3l+2, 3l+3$ in the original representation can therefore be projected onto a macro-site of two spins, with basis states 
\begin{equation}
|00\rangle_{l} 
 \,,\quad |01 \rangle_{l}
  \,,\quad |10\rangle_{l}
   \,,\quad |11\rangle_{l}\,.
\end{equation}
Here, $0$ and $1$ correspond to the two eigenvalues of $Z_{3l+2}\,, Z_{3l+3}$. For a given choice of the eigenvalues, the state on site $3l+1$ is fixed to the eigenstate of $X_{3l+1}$ such that $X_{3l+1} Z_{3l+2} Z_{3l+3}=1$. 

Then, on this reduced space, the action of the generators reads 
\begin{subequations}  
\begin{align}
h_{3p} &= (ZX')_{p-1} \,,\\ 
h_{3p+1} &= (ZZ')_{p-1} (ZZ')_p \,,\\ 
h_{3p+2} &= (Z')_{p-1} (X)_{p} \,, 
\label{eq:reducedrep}
\end{align} 
\end{subequations}  
where on each macro-site $X,Z$, and $X',Z'$ refer respectively to the action on the first or second spin.
Depending on the value of $M({\rm mod}~3)$, the mapping \eqref{eq:reducedrep} may need to be adapted for the rightmost operators $h_m$, but we will not discuss this in detail here.
In summary, the projection has allowed us to map the model onto a chain of $\sim 2M/3$ spins, with the same energy spectrum as the original model but less degeneracies. 

Going further, we could ``factor out'' the rest of the symmetry algebra by fixing the eigenvalues of the remaining generators $\hh_{6l+2}\hh_{6l+3}$. The resulting model has a Hilbert space of dimension $\sim 2^{M/2}$ and we expect that it coincides with the two-chain model presented in Ref~\cite{fendley2019free} and worked out in more detail in Ref. ~\cite{fukai2025quantum} (in that reference, ``minimal representations" of the FFD algebra were presented for any value of $M$). Since this direction is beyond the scope of this work, we will not pursue it further, but it remains an interesting question for future work.

\section{Ancillary free fermions in disguise}
\label{sec:ancillary_fermions}

The $\{\hh_m\}$ operators account for part of the degeneracies ($\sim 2^{M/2}$), observed in the Hamiltonian spectrum. In the following, we describe the ancillary fermionic algebra $\mathcal{A}_{F^\prime}$ accounting for the remaining $\sim 2^{M/6}$ degeneracies. In contrast to $\mathcal{A}_{\widetilde{D}}$, the definition of the algebra  $\mathcal{A}_{F^\prime}$ does not depend on the representation used to describe the FFD, although it depends on the Hamiltonian couplings $\{b_m\}$. We could define $\mathcal{A}_{F'}$ in a concise way by saying that it is the set of all operators written in terms of the $\{h_m\}_{m=1,\ldots M}$ that commute with all fermion creation/annihilation operators. Because of Eq. \eqref{eq:psikdef}, an immediate corollary is that $\mathcal{A}_{F^\prime}$ is generated by all operators that commute with the FFD Hamiltonian and higher conserved charges, as well as with the boundary mode $\chi$. Put differently, $\mathcal{A}_{F^\prime}$ is generated by all operators written in terms of $\{h_m\}_{m=1,\ldots M-1}$ that commute with the Hamiltonian and higher charges: in this last formulation, the absence of $h_M$ is indeed a necessary and sufficient condition for commutation with $\chi$.

\subsection{The (super)symmetry algebra}

The construction of the algebra $\mathcal{A}_{F^\prime}$ follows closely that of the extended SUSY algebra described in Ref.~\cite{fendley2024free}, up to subtle differences which we will highlight below. 

In addition to the generators $h_m$, we need to introduce another boundary mode $\chi'$, squaring to one and commuting with all $\{h_m\}_{m\neq 2}$, while anti-commuting with $h_2$. In the representation Ref.~\eqref{representation}, a simple choice is $\chi'=Z_2$. Next, we define the family of ``fermionic'' operators 
\be
\chi_m =  a_m  ~ \chi' \prod_{i=1}^{m-1} h_i    \,, \qquad m=1, \ldots M+1\,,
\label{eq:chimdef}
\ee 
where $a_m=\frac{b_m b_{m+2} \ldots}{b_{m+1} b_{m+3} \ldots }$, with the dots indicating products running until the end of the chain, that is $M$ or $M-1$ depending on the parity of $m$. The reason for the presence of $\chi'$ is that $\chi_m$ defined as such commutes with all $\{h_m\}_{m=1,\ldots M}$ but two (namely $h_{m-2}$ and $h_{m+1}$), making it very easy to write its commutation relations with $H$, and therefore to find linear combinations of the $\chi_m$ that commute with it.

The $\chi_m$ obey the set of relations
\begin{subequations}    
\begin{align}
(\chi_m)^2 &= (a_m)^2\,,  \\
[\chi_m,\chi_{m+1}]&=0\,,  \\
\{\chi_m,\chi_n\}&=0\,, \qquad |m-n|>1\,.
\end{align}
\end{subequations}
The same relations appeared in Ref.~\cite{fendley2019free}, albeit for different operators, as we now discuss. By expanding Eq.~\eqref{eq:chimdef} in terms of Pauli matrices, we see that the operators $\chi_m$ defined here contain Pauli strings of the form $XYXX\ldots$. This is different from the definition of the operators denoted by $\chi_m$ in Ref.~\cite{fendley2019free}, which contain Pauli strings of the form $XXX\ldots$. This difference has crucial implications when studying the commutation relations with the physical fermions, as we will see shortly. 


Following Ref.~\cite{fendley2019free}, it is not hard to check that the linear combinations
\begin{subequations}    
\begin{align}
O = \sum_{j=0}^{\lfloor \tfrac{M}{2} \rfloor} \chi_{2j+1} \equiv \sum_{o}\chi_o  \,,\\
 E = \sum_{j=1}^{\lfloor \tfrac{M+1}{2} \rfloor} \chi_{2j} \equiv  \sum_{e}\chi_e \,,
 \label{OEdef}
\end{align}
\end{subequations}
commute with $H$, and more generally with all the higher conserved charges. In both definitions the last equality introduces a notation which we will use repeatedly in the following, namely indices associated with the letter $o$ (resp. $e$) are to be summed over all odd (resp. even) integers between $1$ and $M+1$. We can also check that $O$ and $E$ square to a multiple of the identity, and that 
\be 
\{O,E\} = 2 H \,.
\ee 

Next, following once again Ref.~\cite{fendley2019free}, we can define a whole hierarchy of operators $O^{(n)}$, $E^{(n)}$ (with $O^{(1)}=O$, $E^{(1)}=E$), all commuting with $H$ and the higher conserved charges. Using the above notation,
\begin{subequations}
\begin{align}
E^{(2n)} &= \sum_{o_1 \ll e_2 \ll o_3 \ll \ldots \ll e_{2n} } \chi_{o_1} \chi_{e_2} \ldots \chi_{e_{2n}} \,,
\\ 
O^{(2n)} &= \sum_{e_1 \ll o_2 \ll e_3 \ll \ldots \ll o_{2n} } \chi_{e_1} \chi_{o_2} \ldots \chi_{o_{2n}}  \,, \\
E^{(2n+1)} &= \sum_{e_1 \ll o_2 \ll e_3 \ll \ldots \ll e_{2n+1} } \chi_{e_1} \chi_{o_2} \ldots \chi_{e_{2n+1}}\,,
\\
O^{(2n+1)} &= \sum_{o_1 \ll e_2 \ll o_3 \ll \ldots \ll o_{2n+1} } \chi_{o_1} \chi_{e_2} \ldots \chi_{o_{2n+1}}  \,,
\label{eq:EOndef}
\end{align}
\end{subequations}
where the $\ll$ indicate two indices that cannot be consecutive integers. While the definitions hold formally for any $n$, it is easy to see that for a given finite $M$ all operators vanish beyond a certain range. 

Having defined the operators $O^{(n)}$, $E^{(n)}$, let us now study their commutation relations with the physical fermions. As is clear from the definition, all the $\chi_m$ commute with the right boundary mode $\chi$, except for the last, $\chi_{M+1}=\chi' h_1 \ldots h_M$. As a result, out of the operators $O^{(n)}$, $E^{(n)}$, all those whose definition does not involve $\chi_{M+1}$ commute with $\chi$, and therefore with all fermion creation/annihilation operators. This corresponds to the $O^{(n)}$ for odd $M$, and to the $E^{(n)}$ for even $M$. 

We have therefore reached our first step in identifying the symmetry algebra $\mathcal{A}_{F^\prime}$: it contains the operators $O^{(n)}$ for $M$ odd, and the operators $E^{(n)}$ for $M$ even. More precisely, as it will be clear from our subsequent analysis, the symmetry algebra $\mathcal{A}_{F^\prime}$ is generated by the $O^{(2n)}$ (resp. $E^{(2n)}$) together with the bilinears in the $O^{(2n+1)}$ (resp. $E^{(2n+1)}$), such that the extra boundary operator $\chi'$ drops out in such operators (note that, since $\chi'$ does not commute with all $\hh_m$ operators, it acts non-trivially in $\mathcal{H}_{\widetilde{D}}$). This picture is different from the situation in Ref.~\cite{fendley2019free}, where the operators $E^{(n)}$ and $O^{(n)}$ all have non-trivial commutation relations with  the fermions.

\subsection{Commuting transfer matrices}
\label{sec:ancillaryTM}

Next, we study the algebra of the $O^{(n)}$ and $E^{(n)}$ generators. 
First, we observe that all $E^{(n)}$ with even (resp. odd) $n$ commute with one another, and similarly for the $O^{(n)}$. This can be checked by explicit calculation, but in the following we will see how it results, more practically, from a set of relations obeyed by generating functions (``ancillary transfer matrices''). 
To this end, we introduce the families of generating functions 
\begin{subequations}  
\begin{align}
\mathcal{E}_m(u) &= 1+ u \sum_{e\leq m} \chi_e + u^2 \sum_{o \ll e \leq m} \chi_o \chi_e  + \ldots
\\
\mathcal{O}_m(u) &= 1 + u \sum_{o\leq m} \chi_o + u^2 \sum_{e \ll o \leq m} \chi_e \chi_o +\ldots 
 \,,  
\end{align}
\end{subequations}  
whose relation with the SUSY generators of the previous section is $\mathcal{E}_{M+1}(u) = \sum_{m\geq 0} u^m E^{(m)}$, $\mathcal{O}_{M+1}(u) = \sum_{m\geq 0} u^m O^{(m)}$.

For any finite $m$, the generating series truncate at finite order. For instance, $\mathcal{E}_0(u)= \mathcal{O}_0(u) = \mathcal{O}_1(u)=1$, $\mathcal{O}_1(u) = \mathcal{O}_2(u)= 1+u\chi_1$, $\mathcal{E}_2(u) = 1+ u \chi_2$, while for larger $m$, we have the following set of recursion relations
\begin{subequations}    
\begin{align}
\mathcal{E}_{2m}(u) &= \mathcal{E}_{2m-2}(u) + u~ \mathcal{O}_{2m-2}(u) \chi_{2m}\,, \\
\mathcal{O}_{2m}(u) &= \mathcal{O}_{2m-2}(u) + u~ \mathcal{E}_{2m-4}(u) \chi_{2m-1}\,, \\
\mathcal{O}_{2m+1}(u) &= \mathcal{O}_{2m-1}(u) + u~ \mathcal{E}_{2m-1}(u) \chi_{2m+1}\,, \\
\mathcal{E}_{2m+1}(u) &= \mathcal{E}_{2m-1}(u) + u~ \mathcal{O}_{2m-3}(u) \chi_{2m}\,.
\end{align}
\label{recursionrelations}
\end{subequations}

Next, we introduce the symmetric and antisymmetric combinations 
\begin{subequations}  
\begin{align}
\mathcal{E}^\pm_m(u) &= \frac{\mathcal{E}_m(u) \pm \mathcal{E}_m(-u)}{2(i)}  \,, \\ 
\mathcal{O}^\pm_m(u) &= \frac{\mathcal{O}_m(u) \pm \mathcal{O}_m(-u)}{2(i)}\,,
\end{align}
\label{eq:TpmSpm}
\end{subequations}
where the $i$ between parentheses is present for the antisymmetric combinations only. These functions obey recursion relations inherited from \eqref{recursionrelations}. The symmetric (resp. antisymmetric) combinations are polynomials in $u$ containing only even (resp. odd) powers, and we expect from the observation made at the beginning of this section that each of the four combinations should form a family of commuting operators. 
To prove this fact, we include these commutations into a more general set of relations obeyed by the generating functions 
\begin{subequations}  
\begin{align}
[\mathcal{E}_m^\pm(u),\mathcal{E}_m^\pm(v)]&=[\mathcal{O}_m^\pm(u),\mathcal{O}_m^\pm(v)]=0 \,, \\
u \{\mathcal{E}_m^+(u) , \mathcal{O}_m^-(v)\} &=
v \{\mathcal{E}_m^+(v) , \mathcal{O}_m^-(u)\} \\
v \{\mathcal{E}_m^-(u) , \mathcal{O}_m^+(v)\} &=
u \{\mathcal{E}_m^-(v) , \mathcal{O}_m^+(u)\}   \,,
\label{algebraancillaryTM}
\end{align}
\end{subequations}  
and which are easy to prove recursively, using Eq.~\eqref{recursionrelations}.

\subsection{The ancillary fermions}

We now have all the ingredients for diagonalizing the operators $O^{(n)}$ and $E^{(n)}$. As we will see, the diagonalization follows very closely Fendley's original solution of the FFD Hamiltonian~\cite{fendley2019free}, and involves another set of ancillary free fermions commuting with the physical ones. 

To see this, we study the products 
\begin{subequations}  
\begin{align}
\mathbb{E}_m^\pm(u) &= \mathcal{E}_m^\pm(\sqrt{u}) \mathcal{E}_m^\pm(\sqrt{-u})  
\,,
\\
\mathbb{O}_m^\pm(u) &= \mathcal{O}_m^\pm(\sqrt{u}) \mathcal{O}_m^\pm(\sqrt{-u})   \,,
\end{align}
\end{subequations}  
which are the analog of the products $T(u)T(-u)$ in Fendley's original construction. 
Those are polynomials in $u$, and obey the recursion relations 
\begin{subequations}    
\begin{align}
\mathbb{E}_{2m}^\pm(u) &=\mathbb{E}_{2m-2}^\pm(u) + u~ \mathbb{O}_{2m-2}^\mp(u) a_{2m}^2 \\
\mathbb{O}_{2m}^\mp(u) &=\mathbb{O}_{2m-2}^\mp(u) + u~ \mathbb{E}_{2m-4}^\pm(u) a_{2m-1}^2 \\
\mathbb{O}_{2m+1}^\pm(u) &=\mathbb{O}_{2m-1}^\pm(u) + u~ \mathbb{E}_{2m-1}^\mp(u) a_{2m+1}^2 \\
\mathbb{E}_{2m+1}^\mp(u) &=\mathbb{E}_{2m-1}^\mp(u) + u~ \mathbb{O}_{2m-3}^\pm(u) a_{2m}^2\,.
\end{align}
\label{eq:recursionrelationpolynomials}
\end{subequations}
Together with the fact that for $m\leq 0$ we have $\mathbb{E}^+_m(u) = \mathbb{O}^+_m(u) = 1 $, $\mathbb{E}^-_m(u) = \mathbb{O}^-_m(u) = 0$, these recursion relations guarantee that, for all $m$, the operators $\mathbb{E}_m^\pm(u)$, $\mathbb{O}_m^\pm(u)$ are proportional to the identity. 

For a given value of $M$, let us focus on the generating function of even-order commuting charges in the algebra $\mathcal{A}_{F^\prime}$, namely $\mathcal{E}_{M}^+(u)=\mathcal{E}_{M+1}^+(u)$ for $M$ even (resp. $\mathcal{O}_{M}^+(u)=\mathcal{O}_{M+1}^+(u)$ for $M$ odd), and on the corresponding generating polynomial $\mathcal{P}_{M}(u) \equiv \mathbb{E}_{M}^+(u)$ for $M$ even (resp. $\mathbb{O}_{M}^+(u)$ for $M$ odd).
Based on the recursion relations $\eqref{eq:recursionrelationpolynomials}$, it is easy to see that the degree of this polynomial is $2S' = 2 \lfloor \frac{M+2}{6} \rfloor$, with $S'$ pairs of opposite roots. Calling the latter $\{\pm u'_k\}_{k=1,\ldots S'} \equiv \{\pm 1/ \epsilon'_k\}_{k=1,\ldots S'}$, it is then straightforward to recast the eigenvalues of the commuting charges $E^{(2n)}$ (resp. $O^{(2n)}$) in terms of the parameters $\epsilon'_k$. In particular, $E^{(2)}$ (resp. $O^{(2)}$) has $2^{S'}$ distinct eigenvalues of the form  
\begin{equation}
\pm \epsilon'_1 \pm \epsilon'_2 \ldots \pm \epsilon'_{S'}\,,    
\end{equation}
where all signs can be chosen independently. These operators can therefore be considered as ancillary free fermionic (albeit non local) ``Hamiltonians'', commuting with the physical one, and which can be used to resolve the remaining degeneracies. 

Going further, we can construct the corresponding fermionic creation and annihilation operators. Since the derivation is similar to those of Ref.~\cite{fendley2019free}, we omit it and only provide the final result: the ancillary creation and annihilation operators read 
\be 
\Psi'_{\pm k} \propto \begin{cases}  \mathcal{E}_M(\sqrt{\pm u_k}) ~E~ \mathcal{E}_M(\sqrt{\mp u_k}) \qquad \mbox{for $M$ even}
\\
 \mathcal{O}_M(\sqrt{\pm u_k}) ~O~ \mathcal{O}_M(\sqrt{\mp u_k}) \qquad \mbox{for $M$ odd}
\end{cases} \,,
\label{ancillaryfermions}
\ee 
for each $k=1\ldots S'$, where the proportionality factors can be fixed such that $\{\Psi'_k, \Psi'_l\} = \delta_{k+l,0}$.

\subsection{The symmetry algebra $\mathcal{A}_{F^\prime}$ in terms of the ancillary fermions}

Before concluding, it is useful to summarize our results. In the previous sections, we found operators $\Psi'_k$ ($k \in \pm \{1,\ldots S'\}$) that obey canonical fermionic commutation relations. By construction, they commute with all the physical fermion creation/annihilation operators. However, because of the $E$ (resp. $O$) factor in Eq.~\eqref{ancillaryfermions}, which involves an operator $\chi'$ anti-commuting with some of the $\widetilde{h}_m$, only bilinear combinations of those fermions commute with the full algebra $\{\hh_m\}$ and therefore have a trivial action in both $\mathcal{H}_F$ and $\mathcal{H}_{\widetilde{D}}$. 

We are therefore ready to state our final proposition, namely that the algebra $\mathcal{A}_{F^\prime}$ is generated by all bilinears in the ancillary fermions $\Psi'_k$. This is confirmed by a counting of dimensions presented in Table \ref{table:summary}, cf. Sec.~\ref{sec:result_overview}, where it is checked that the ancillary fermionic modes, together with the algebra $\mathcal{A}_{\widetilde{D}}$ of Sec.~\ref{sec:tilde_h}, indeed accounts for all the degeneracies observed in the spectrum.

\subsection{The remaining zero mode for $M=6k+3$}

As mentioned above, for the special case $M=6k+3$, the algebras $\mathcal{A}_{\widetilde{D}}$ and $\mathcal{A}_{F'}$ are enough to lift all degeneracy factors of the spectrum, but one. 
This can be explained by the existence of an additional zero mode, which commutes both with the physical fermions and with $\mathcal{A}_{\widetilde{D}}$ and $\mathcal{A}_{F'}$. 
For $M=3$, such an operator can easily be found as $(h_1+h_2)\chi_M$. For $M=9$, it can be found as a bilinear in the $\{h_i\}_{i\leq M-1}$ multiplied by $\chi_M$. For general $M=6k+3$ we conjecture that an operator of such a form should exist, namely some multilinear combination of $\{h_i\}_{i\leq M-1}$ multiplied by $\chi_M$. 
Furthermore, it should be possible to relate the existence of this operator to the zero modes evidenced in \cite{vona2014exact} more generally for all $M\notin 3\mathbb{N}+1$. However, a difference between our approach and the zero modes of \cite{vona2014exact} is that the latter do not necessarily commute with the rest of the symmetry algebra, that is, with $\mathcal{A}_{\widetilde{D}}$ and $\mathcal{A}_{F'}$.

\section{Outlook}
\label{sec:outlook}
FFD models have a relative short history, especially compared to the well understood JW solvable models, and still present many challenges. In this work, we have focused on one particular aspect which has so far received little attention: the characterization of the FFD Hilbert-space structure. We have shown that the latter admits an exact factorization into free-fermionic and degenerate subspaces, respectively. By constructing spin operators generating the operator algebra supported on the degenerate subspace, we were able to fully resolve all the FFD Hamiltonian degeneracies.

Our work opens several natural directions. Most naturally, our results provide new tools for the computation of correlations functions, both in an out of equilibrium. Previously, this problem was tackled by seeking a solution to the so-called ``inverse problem'', consisting in expressing local spin operators in terms of fermionic ones. In particular Ref.~\cite{vona2014exact} found a few special local spin operators that admit a simple representation in terms of the free fermionic operators, making it possible to study their dynamics explicitly. In our work, we have completely characterized the set of operators generating the symmetry algebra, which is supported on the degenerate subspace and, by definition, not evolving under the FFD Hamiltonian dynamics. This allows us go further, and consider spin operators admitting a simple representation not only in terms of the fermionic ones, but also of the elements of the symmetry algebra. Finally, another natural direction pertains to extending our constructions to different FFD models, including the one presented in Ref.~\cite{fendley2024free}, the parafermionic generalizations developed in Refs.~\cite{alcaraz2020free,alcaraz2020integrable}, the FFD quantum circuits studied in Ref.~\cite{fukai2025quantum}  and more generally the families FFD-solvable models characterized by their frustration graphs~\cite{fukai2025free,chapman2023unified}. The work \cite{chapman2023unified}, in particular, introduces a further fragmentation of the Hilbert space  in terms of cycle symmetries: it would be interesting to understand how this relates to the Hilbert space factorization presented in this work. These directions will be explored in future research.

\section*{Acknowledgments}
We thank Paul Fendley for discussions, and Dávid Szász-Schagrin and Daniele Cristani for collaboration on related topics. EV acknowledges support from the CNRS-IEA. The work of LP was funded by the European Union (ERC, QUANTHEM, 101114881). Views and opinions expressed are however those of the author(s) only and do not necessarily reflect those of the European Union or the European Research Council Executive Agency. Neither the European Union nor the granting authority can be held responsible for them.

\bibliography{bibliography}

\end{document}